\documentclass[twocolumn,english,aps,prb,floatfix,preprintnumbers,showpacs,amsfonts,amssymb,superscriptaddress]{revtex4}
\usepackage{ae,aecompl}
\usepackage[T1]{fontenc}
\usepackage[latin1]{inputenc}
\usepackage{amsmath}
\usepackage{graphicx}
\usepackage{amssymb}

\makeatletter

\providecommand{\tabularnewline}{\\}

\@ifundefined{textcolor}{}
{%
 \definecolor{BLACK}{gray}{0}
 \definecolor{WHITE}{gray}{1}
 \definecolor{RED}{rgb}{1,0,0}
 \definecolor{GREEN}{rgb}{0,1,0}
 \definecolor{BLUE}{rgb}{0,0,1}
 \definecolor{CYAN}{cmyk}{1,0,0,0}
 \definecolor{MAGENTA}{cmyk}{0,1,0,0}
 \definecolor{YELLOW}{cmyk}{0,0,1,0}
 }

\usepackage{amscd}
\usepackage{bm}
\usepackage{graphics,psfrag}
\usepackage{graphicx,psfrag}

\makeatother

\usepackage{babel}

\begin{document}

\title{Finite-size corrections of the Entanglement Entropy of critical quantum
chains }

\author{J.~C.~Xavier}

\affiliation{Instituto de F\'{\i}sica, Universidade Federal de Uberlândia, Caixa
Postal 593, 38400-902 Uberlândia, MG, Brazil }

\affiliation{Instituto de F\'{\i}sica de São Carlos, Universidade de São Paulo,
Caixa Postal 369, 13560-970 São Carlos, SP, Brazil}

\author{F.~C.~Alcaraz}

\affiliation{Instituto de F\'{\i}sica de São Carlos, Universidade de São Paulo,
Caixa Postal 369, 13560-970 São Carlos, SP, Brazil}

\date{\today{}}
\begin{abstract}
Using the density matrix renormalization group, we calculated the
finite-size corrections of the entanglement  $\alpha$-Rényi entropy
of a single interval for several critical quantum chains. We considered models with $U(1)$
symmetry like the spin-1/2 XXZ and spin-1 Fateev-Zamolodchikov models,
as well as models with discrete symmetries such as the Ising, the Blume-Capel,
and the three-state Potts models. These corrections contain physically
relevant information. Their amplitudes, that depend on the value of
$\alpha$, are related to the dimensions of operators in the conformal
field theory governing the long-distance correlations of the critical
quantum chains. The obtained results together with
earlier exact and numerical ones allow us to formulate some general
conjectures about the operator responsible for the leading finite-size
correction of the $\alpha$-Rényi entropies. We conjecture that the
exponent of the leading finite-size correction of the $\alpha$-Rényi
entropies is $p_{\alpha}=2X_{\epsilon}/\alpha$ for $\alpha>1$ and
$p_{1}=\nu$, where $X_{\epsilon}$ denotes 
the dimensions of the energy operator of the model and $\nu=2$ for all 
the models.
\end{abstract}


\maketitle

\section{Introduction}

Usually, we are interested in  ground state properties
of quantum critical chains in the thermodynamic limit. However,
obtaining the physical properties in this limit, in general, is not
possible. On the other hand, many numerical techniques 
that give accurate information
about the system, work for finite systems. It is therefore of fundamental
importance to obtain the physical properties of the infinite system of 
interest from its formulation in a \emph{finite-size} geometry.
 Indeed, various physical quantities of interest are directly
related to the finite-size scaling corrections of some physical measure.
In this work, we are interested in the scaling corrections of the
entanglement entropy of one-dimensional critical systems, which has
been intensely debated in recent years. 

Consider a quantum chain with $L$ sites, described by a pure state
whose density operator is $\rho$. Let us consider that the system
is composed by the subsystems ${\cal A}$ with $\ell$ sites ($\ell=1,\ldots,L$)
and ${\cal B}$ with $L-\ell$ sites. The Rényi entropy is defined
as

\begin{equation}
S_{\alpha}(L,\ell)=\frac{1}{1-\alpha}\ln Tr(\rho_{{\cal {A}}}^{\alpha}),\label{eq:renyientropy}\end{equation}
where  $\rho_{{\cal {A}}}=\mbox{Tr}_{{\cal {B}}}\rho$ is the reduced
density matrix of the subsystem ${\cal A}$. The von Neumann entropy,
also known as entanglement entropy, is the particular limit  $\alpha \to 1$.

In the scaling regime $1<<\ell<<L$, it is expected that for the critical
one-dimensional systems, under periodic boundary conditions (PBC),
the Rényi entropy of the ground state behaves as

\begin{equation}
S_{\alpha}(L,\ell)=S_{\alpha}^{CFT}(L,\ell)+S_{\alpha}^{UCS}(L,\ell).\label{eq:entropyb}\end{equation}
 The first term in this equation, is the conformal field theory (CFT)
prediction and is given by \citealp{cold,cardyentan,entroreviewcalabrese,affleckboundary,prlkorepin} 

\begin{equation}
S_{\alpha}^{CFT}=\frac{c}{6}\left(1+\frac{1}{\alpha}\right)\ln\left[\frac{L}{\pi}\sin\left(\frac{\pi\ell}{L}\right)\right]+d_{\alpha},\label{eq:entropyCFT}\end{equation}
 where $c$ is the central charge and $d_{\alpha}$ is a non-universal
constant.

Laflorencie \emph{et al} in Ref. \onlinecite{entropyaffleckosc}
were the first to notice, in an investigation of the spin-1/2 Heisenberg
chain under open boundary condition, that unusual corrections to scaling
appear in the Rényi entropy $S_{\alpha}(L,\ell)$ with $\alpha=1$.
They noted that the standard CFT term {[}Eq. \ref{eq:entropyCFT}{]}
could not explain a strong oscillation observed in the von Neumann
entropy, which seems to have origin its in the oscillating behavior of 
the spin-spin correlations  
of the Hamiltonian. These strong oscillations, for the quantum open
chains, were observed subsequently by several authors.\citealp{entropyaffleckosc,total,xavierentanglement,entropyosc,xxPBCh,calabreseOBC,calabreserandom}
More recently, Calabrese \emph{et al}. in Ref. \onlinecite{entropyosc}
(see also Ref. \onlinecite{xxPBCh}) investigated the anisotropic
spin-1/2 Heisenberg chain with PBC and they noted that although 
those oscillations are not present in the von Neumann entropy $S_{1}$,
they are still present for the Rényi index $\alpha>$1. Based on exact
results and also in numerical calculations of the spin-1/2 XXZ chain
with PBC and zero magnetic filed, Calabrese and collaborators conjectured
that $S_{\alpha}^{UCS}$, apart from a non-universal constant $g_{\alpha}$,
has the following universal behavior\citealp{entropyosc,xxPBCh,calabreseOBC}(see
also Ref. \onlinecite{entropyaffleckosc}) \begin{eqnarray}
 &  & S_{\alpha}^{UCS}=D_{\alpha}-d_{\alpha}=g_{\alpha}\cos(\kappa\ell)\left[\sin\left(\frac{\pi\ell}{L}\right)\right]^{-p_{\alpha}},\label{eq:entropyUnusual}\end{eqnarray}
where for convenience we introduce the function $D_{\alpha}(L,\ell)$.
In (4) $p_{\alpha}$ is a critical exponent and $\kappa$, which has
distint values for different models, gives the spatial period
$\lambda=\frac{2\pi}{\kappa}$ of the oscillations.\citep{commphi}
For instance, for the spin-$s$ $XXZ$ chains at zero magnetic field
$\kappa_{XXZ}=\pi$. For these chains, the universal exponent is $p_{\alpha}=\frac{2K}{\alpha}$,
where $K$ is the Luttinger liquid parameter of the underlying CFT.\citealp{entropyosc,xavieralcarazosc}
For the Ising model $\kappa_{Ising}=2\kappa_{XXZ}$, and we see no
oscillations.The origin of the oscillating factor {[}$\cos(\kappa\ell)${]}
was not yet completely understood, however it has been observed in
the systems whose spin correlations show an oscillating behavior.

By now, the observation of the unusual corrections to scaling in entanglement
entropy with the predicted exponents {[}Eq. (\ref{eq:entropyUnusual}){]}
were confirmed only for: the spin-$s$ XXZ chains,\citealp{entropyosc,xavieralcarazosc}
the 1D attractive Hubbard model, and in a dipolar boson quantum chain.\citealp{taddiaosc}
As observed by Cardy and Calabrese in Ref.~\onlinecite{cardyosc}
the origin of the exponent $p_{\alpha}$ in Eq.~ (\ref{eq:entropyUnusual})
is the conical singularities produced by the conformal mapping used
to describe the reduced density matrix $\rho_{{\cal A}}=\mbox{Tr}_{{\cal B}}\rho$
in CFT. This exponent is related to the scaling dimension $X^{con}$
of an operator of the underlying CFT by $p_{\alpha}=2X^{con}/\alpha$.\citealp{cardyosc}. This operator can be relevant ($X^{con}<2$) or not ($X^{con} \geq 2$).
Beyond these corrections, we may also have the usual ones coming from the
irrelevant operators (dimension $X^{I}>2$), whose leading contributions
to (\ref{eq:entropyUnusual}) are of order ${\ell}^{-2(X^{I}-2)}$,
for any $\alpha$.\citealp{cardyosc} 

Recently, Calabrese and Essler in Ref. \onlinecite{xxPBCh}, based
on an exact calculation for the XX quantum chain, found the appearance
of a correction term of order ${\ell}^{-2}$ for any value of $\alpha$.
As discussed in Ref.~\onlinecite{cardyosc}, we should expect $\alpha$-independent
correction terms from the irrelevant operators. This would imply that
the contributing irrelevant operator has dimension $X^{I}=3$. Although
such an irrelevant operator exists in the conformal tower of the XX model,
it is quite unlikely that this contribution comes from such an operator.
In this model, the operator responsible for the leading finite-size
corrections of the eigenenergies has dimension $X^{I}=4$, \citealp{chico1}
which  would produce a correction term of order ${\ell}^{-4}$. A possible
explanation is that the conical singularities may also produce an $\alpha$-independent
contribution to the corrections to scaling, besides the term of order
${\ell}^{-2X^{con}/{\alpha}}$. Although not completely clear, such
contribution might come from the combination of contributions of
the relevant operators,\citealp{priveEssler} and in the case of the
XX chain it ends up with the main contribution of order ${\ell}^{-\nu}$,
with $\nu=2$.

Although the CFT analysis predicts the possible finite-size corrections
of the Rényi entropies, several general questions remain to be answered
for a better understanding: a) Among the possible relevant operators
in the CFT what should be the leading one, with dimension $X^{con}$,
responsible for the corrections coming from the conical singularities
giving $\alpha$-dependent contributions of order ${\ell}^{-2X^{con}/\alpha}$
? b) What should be the leading irrelevant operator with dimension
$X^{I}$, giving the contribution $O({\ell}^{-2(X^{I}-2)})$ ? c) Are the $\alpha$-independent
corrections ${\ell}^{-\nu}$, coming from the conical singularities, as happened in
the XX case, general ? In this case what is the value of $\nu$ ? 
 d) The
oscillating behavior observed in the $\alpha>1$ Rényi entropy is
absent in the quantum Ising chain. Should we obtain the oscillatory
behavior of the entropies for other quantum chains with $U(1)$ symmetries,
like the XXZ chain?

In order to test the above predictions and answer the above questions,
we present in this paper the calculation of the corrections to scaling
in the entropies of several critical quantum chains belonging to distinct
universality classes of critical behavior. We consider models with
$U(1)$, $Z(2)$ and $Z(3)$ symmetries. The models with U(1) symmetry
are the spin-1/2 XXZ\textbf{ }quantum chain and the spin-1 Fateev-Zamolodchikov
model.\citealp{spin1ZM} The models with $Z(2)$ and $Z(3)$ symmetries
we consider are the quantum Blume-Capel and the quantum 3-state
Potts model, respectively.

\section{RESULTS}

%In order to study the corrections to scaling of the Rényi entropies
%for general critical models it is convenient to define the difference \begin{eqnarray}
% &  & D_{\alpha}(L,{\ell})=S_{\alpha}(L,{\ell})-\frac{c}{6}\left(1+\frac{1}{\alpha}\right)\ln\left[\frac{L}{\pi}\sin\left(\frac{\pi\ell}{L}\right)\right].\label{eq:diffentropy}\end{eqnarray}
 From the discussions presented in the introduction we expect, in
the region where the system size $L$ and subsystem size ${\ell}$
are large ($L>>{\ell}>>1$), the general behavior: \begin{eqnarray}
 &  & D_{\alpha}(L,{\ell})=d_{\alpha}+g_{\alpha}\cos(\kappa{\ell})\left[\sin\left(\frac{\pi\ell}{L}\right)\right]^{-2X^{con}/{\alpha}}\nonumber \\
 &  & +a_{\alpha}\left[\sin\left(\frac{\pi\ell}{L}\right)\right]^{-\nu }+b_{\alpha}\left[\sin\left(\frac{\pi\ell}{L}\right)\right]^{-2(X^{I}-2)}+\cdots.\label{eq:diffentropyb}\end{eqnarray}
 The second and third terms are the leading contributions 
%coming from
%the relevant operators with dimension $X_{R}$ and $\bar{X}_{R}$
due to the conical singularities,\citealp{cardyosc} and the last
term is the leading contribution due to the standard correction to
scaling operator, with dimension $X^{I}$. 
The operator that produces the second term in (6) has dimension $X^{con}$, 
while the relation of the $\alpha$-independent exponent $\nu$ with the 
dimension of operators is unknown.  
Models exhibiting oscillations
in the $\alpha$-entropies have $g_{\alpha}\neq0$ and $\kappa\neq0$
($\mbox{mod}\;2\pi$). Note that only the exponent of the second leading
corrections depend on the value of $\alpha$. We intend to evaluate
the power of the dominant term in (\ref{eq:diffentropyb}), which
we denote by $p_{\alpha}$. In order to estimate the exponent $p_{\alpha},$
we fit our data with the following equation

\begin{equation}
D_{\alpha}(L,{\ell})=d_{\alpha}+f_{\alpha}\left[\cos(\kappa\ell)\right]^{(1-\delta_{\alpha,1})}\left[\sin\left(\frac{\pi\ell}{L}\right)\right]^{-p_{\alpha}}.\label{eq:fit}\end{equation}

We report first the results already known from earlier studies for
some models. The spin-1/2 XXZ chain shows oscillations for $\alpha>1$.
These corrections come from the operator with dimension $X^{con}=K$,
where $K$ is the Luttinger liquid parameter.\citealp{entropyosc}
This is the scaling dimension of the \textit{energy operator} of the
model.\cite{commCFT}
For $\alpha=1$ the oscillations do not appear, i. e., $g_{1}=0$,
and the dominant term in (\ref{eq:diffentropyb}) \emph{is not known}
for general values of the anisotropy of the model. In the particular
case where we have the XX model (free fermion case), the dominant
correction for $D_{1}(L,{\ell})$ is given by the 
third term in (6) with $\nu=2$.
\citealp{xxPBCh} 

The non integrable spin-$s$ XXZ chain ($s=1,3/2,\ldots$) on its
critical regions also shows entropy oscillations for $\alpha>1$,
but the amplitudes $g_{\alpha}$ decreases strongly as we increase
the value of the spin $s$.\citealp{xavieralcarazosc} On these models,
the amplitudes of the entropy oscillations are also ruled by the \textit{energy
operator} of the model, with dimension $X_{\epsilon}=X^{con}=K$, given
by the Luttinger parameter of the underlying CFT.

The results for the Ising model can be obtained from those of the
spin-1/2 XX model due to the exact correspondence of their entropies,
as shown in Ref. \onlinecite{igloixy}. The Ising model does not show 
oscillations,
for any $\alpha$,  due to the fact that $\kappa_{Ising}=2\kappa_{XXZ}=2\pi$.
We also have $g_{1}=0$ and $g_{\alpha}\neq0$ ($\alpha>1$) with
the dominant correction term given by ${X}^{con}=1$, which is also
the dimension of the \textit{energy operator} of the model. In the
case $\alpha=1$ the dominant correction comes from the 
 third term of (5) with $\nu=2$.

Below, we present new numerical results for three models: (i) the
Blume-Capel Model (BCM), (ii) three-state Potts Model (3SPM), and
(iii) the Fateev-Zamolodchikov quantum chain (FZQC). We investigated these
models with the density-matrix renormalization group (DMRG) technique\citealp{white}
under PBC. We have done 4\textendash{}8 sweeps. For the BCM and the
3SPM we used typically $m=400$ states per block. This number of states
kept in the truncation process is enough to give very precise results,
the discarded weight being typically about $10^{-12}$. However, for
the FZQC we used a much larger number of states in order to obtain
 precise results (up to $m=3000$). In this case, the discarded
weight was typically $10^{-8}$ in the final sweep.

\subsection{The spin-1 Blume-Capel model}

We begin by introducing the spin-1 BCM quantum chain. This model is
obtained by the time-continuum limit of the well known BCM in two
dimensions.\citealp{PhysRevB.32.7469} It describes the dynamics of
spin-1 localized particles, with Hamiltonian given by

\begin{equation}
H_{BC}=-\sum_{j}\left(s_{j}^{z}s_{j+1}^{z}-\delta(s_{j}^{z})^{2}-\gamma s_{j}^{x}\right),\label{eq:BC}\end{equation}
 where $s^{x}$ and $s^{z}$ are the spin-1 $SU(2)$ operators. The
phase diagram, in the $\delta-\gamma$ plane, is known from earlier
numerical finite-size scaling studies based in the crossing of the
mass gap energies (see Fig. 1 of Ref. \onlinecite{PhysRevB.32.7469}).
For values $\gamma>\gamma_{\mbox{\scriptsize tr}}$ the Hamiltonian
has a quantum critical line $\delta_{c}(\gamma)$ governed by a CFT
in the same universality class of the quantum Ising chain, i. e.,
central charge $c=1/2$. At $\gamma_{\mbox{\scriptsize tr}}$ the
model has a quantum tricritical point at $\delta_{\mbox{\scriptsize tr}}$
in the universality class of the tricritical Ising model, having central
charge $c=7/10$. For $\gamma<\gamma_{\mbox{\scriptsize tr}}$ there
is a line $\delta=\delta_{gap}(\gamma)$ of first-order phase transitions.
Recently, highly accurate estimates of some points of the critical
line were obtained by a new method based on the entanglement entropy.\citep{xavieralcarazQCP}
In particular, the tricritical point was located at $\gamma_{\mbox{tr}}=0.41563$
and $\delta_{\mbox{tr}}=0.91024$. Since the model has a critical
line in the universality class of the Ising model (IM) it is  interesting,
for the sake of comparison and as benchmark tests, to consider also
the integrable IM. The Hamiltonian of the IM is given by

\begin{equation}
H_{Ising}=-\sum_{j}\left(\sigma_{j}^{x}\sigma_{j+1}^{x}+\lambda\sigma_{j}^{z}\right),\label{eq:ising}\end{equation}
 where $\sigma^{x}$, $\sigma^{z}$ are spin-1/2 Pauli matrices. This
model has a critical point $\lambda_{c}=1$ that can be obtained from
its exact solution, or even more simply from its self dual property.\citep{Kogut1}

\begin{figure}
\psfrag{ell}{$\ell$}
\begin{centering}
%\psfrag{ell}{$\ell$}
\includegraphics[scale=0.24]{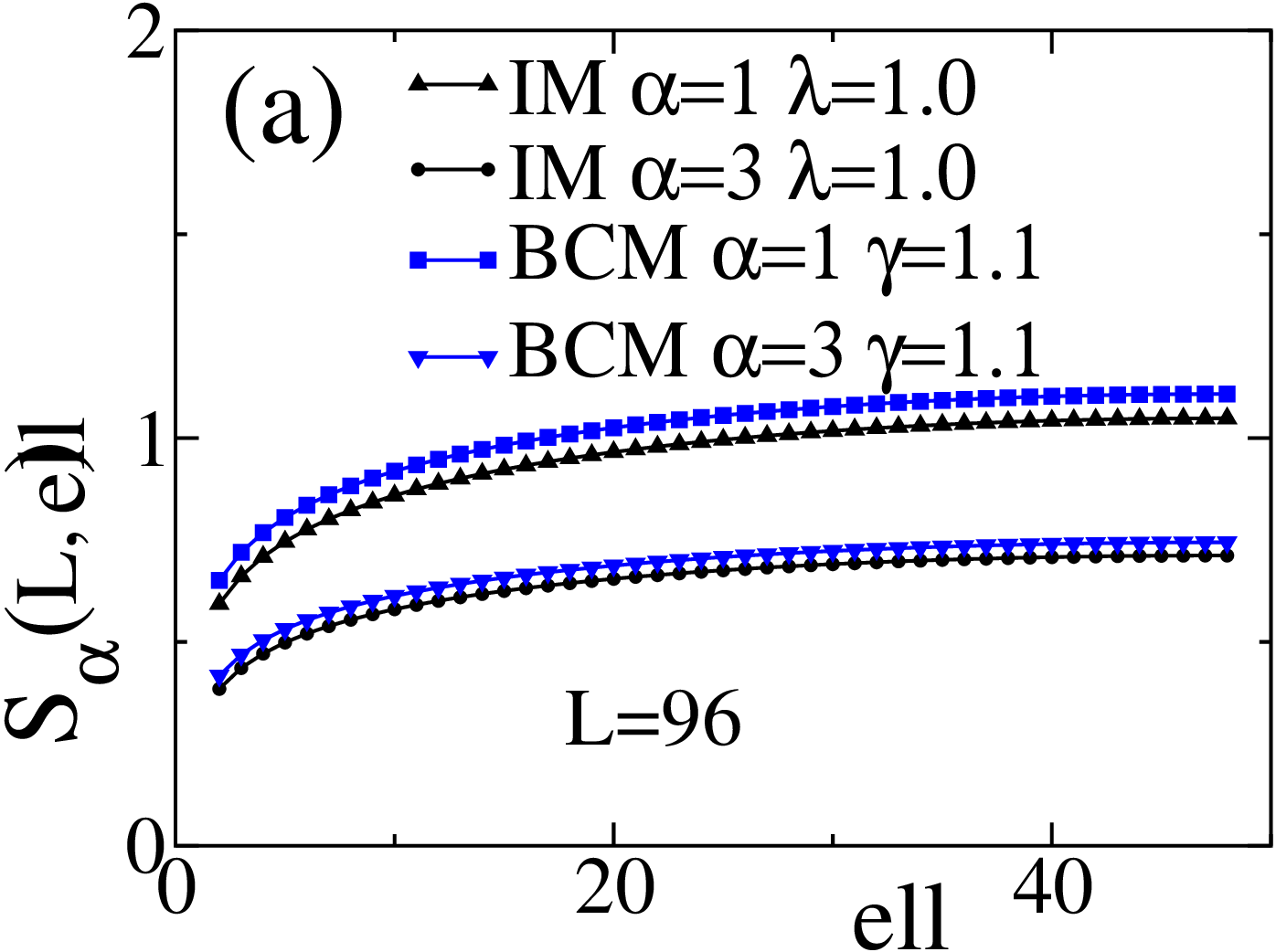}
\par\end{centering}

\begin{centering}
\includegraphics[scale=0.24]{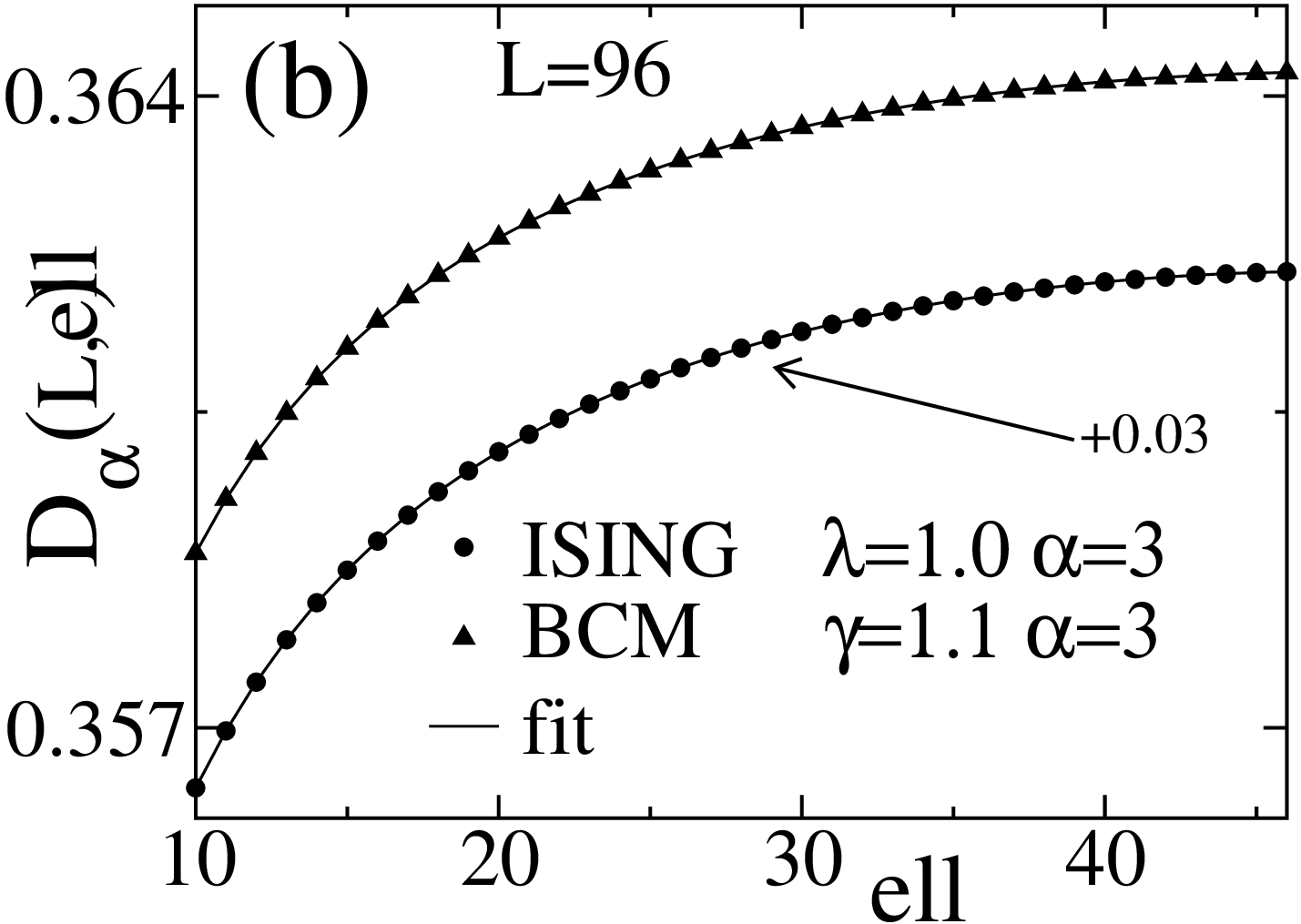}
\par\end{centering}

%\begin{centering}
%\includegraphics[scale=0.24]{fig1c}
%\par\end{centering}

\caption{\label{fig1} (Color online). Results of the Rényi entropy $S_{\alpha}(L,\ell)$
and of the function $D_{\alpha}(L,\ell)$ for the Ising and Blume-Capel
models with PBC and size $L=96$. (a) $S_{\alpha}(L,\ell)$ vs $\ell$
for two values of $\alpha$, and some values of coupling (see legend).
The symbols are the numerical data and the solid lines connect the
fitted points using Eq. (\ref{eq:entropyCFT}). (b) $D_{\alpha}(L,\ell)$
vs $\ell$ for $\alpha=3$. We added 0.03 in the values of the Ising
entropy in order to see better both data in the same figure. 
%(c) Comparison
%of $S_{\alpha}(L,\ell)$ for the BCM at a critical $(\gamma_{c},\delta_{c})$
%{[}tricitical $(\gamma_{tr},\delta_{tri})${]} point and at close
%point $(\gamma_{c},\delta_{c}+\epsilon)${[}$(\gamma_{tr}+\epsilon,\delta_{tri}+\epsilon)${]}
%with $\epsilon=0.5\times10^{-4}$. Only few sites are presented.
}

\end{figure}

In Fig. \ref{fig1}(a), we show the Rényi entropy $S_{\alpha}(L,\ell)$,
as a function of $\ell$, for the IM and the BCM with PBC and lattice
size $L=96$. In the IM case, we select the exact critical point $\lambda=1$.
Since in the BCM we do not know the exact values of the critical points,
we used the best estimate of the critical points available from previous
work.\citep{xavieralcarazQCP} In particular, for $\gamma=1.1$ we
used $\delta_{c}=0.31357$, where we expect an error in the last digit
(see Ref. \onlinecite{xavieralcarazQCP}). In both cases, the fits
to Eq.~(\ref{eq:entropyCFT}) give a central charge value very close
to the exact one. We found that $|c^{fit}-c^{exact}|\lesssim10^{-3}$
for all critical couplings considered (some of these estimates are
presented in Table \ref{tab:BCMpalpha}). We emphasize that 
the discrepancies between 
the exact values and the numerical
data are due to the finite-size of the systems considered.
The numerical errors (inferred from the dependence of the entropy with $m$) 
are smaller than the size of     
the symbols  in the figures. For the models with
discrete (continuous) symmetry the errors are smaller than $10^{-5}$ 
($10^{-3}$).

The results presented in Fig. \ref{fig1}(a) are nice illustrations
of the universal behavior of the Rényi entropies at the critical points.
Although the entropies are calculated from ground states defined in
Hilbert spaces of quite distinct dimensions ($2^{L}$ and $3^{L}$)
the entropies show the same behavior. The small deviation shown in
this figure is due to the non-universal constant $d_{\alpha}$ in
(\ref{eq:entropyCFT}). Notice also that  both models, 
in the Ising Universality class, show no oscillations in the entropies
(even in the OBC case, not shown), i. e., $\kappa=0$ in (\ref{eq:diffentropyb}). 

The corrections to scaling contributions are calculated from the behavior
of $D_{\alpha}(L,\ell)$ given in (\ref{eq:diffentropyb}). As illustrative
examples, let us consider, as in Fig. 1(a), the IM at the critical
coupling $\lambda_{c}=1$ and the BCM at the couplings $\gamma=1.1$
and $\delta_{c}=0.31357$. In Fig. \ref{fig1}(b), we present $D_{\alpha}(L,\ell)$
for the critical couplings of the quantum chains with PBC, $\alpha=3$,
and lattice size $L=96$ (similar results are found for $L=64$).
Since at these couplings, both models have central charge $c=1/2$
we set this value in Eq.  (\ref{eq:entropyCFT}). Note
in these two figures that the changes in $D_{\alpha}(L,\ell)$, as
function of $\ell$, are quite small, which indicate that the non-universal
constants $g_{\alpha}$, $a_{\alpha}$ and $b_{\alpha}$ that appear
in (\ref{eq:diffentropyb}) are very small. This fact makes the task
of determining the exponent $p_{\alpha}$ in (\ref{eq:fit}) a challenge.
Since we are interested in the asymptotic behavior of $D_{\alpha}(L,\ell)$,
we have discarded the first 10 values of $\ell$. Comparison of the
entropies with different values of $m$ indicate that the errors in
the DMRG evaluations of the entropies are around $\sim10^{-6}-10^{-7}$.
Due to this fact, we also discarded the sites $\ell$ at which $|D_{\alpha}(L,\ell)-D_{\alpha}(L,\ell+1)|<10^{-5}$. 

 We get $p_{\alpha}$ by fitting our data to Eq. (\ref{eq:fit}).
The solid lines, in Fig. \ref{fig1}(b), connect the fitted points.
In order to check if we are able to extract good estimates of $p_{\alpha}$
with the above mentioned procedure, we compare the values of  $p_{\alpha}$,
acquired with the fit procedure with the exact ones ($p_{\alpha}=\frac{2}{\alpha}$)
for the Ising model. The finite-size estimates for the exponents $p_{2},$
$p_{3}$ and $p_{4}$ are depicted in Table \ref{tab:Isingpalpha}.
As we can observe in this table, the finite-size estimates differ very little
(less than $5\times10^{-2}$) from the exacts values. We then believe that this
approach will also  provide good estimates of the exponents $p_{\alpha}$
for other models like the BCM.

Before presenting our results for the BCM, it is interesting to mention
that, for the IM, we got \emph{only} 8 points of $D_{\alpha=1}^{Ising}(L,\ell)$
satisfying $|D_{1}(L,\ell)-D_{1}(L,\ell+1)|<10^{-5}$ for $\ell>2$.
This in fact is expected, since for $\alpha=1$ in the IM, the dominant
correction to scaling is conjectured as
$D_{1}(L,\ell)=d_{1}+\frac{1}{60}\frac{\pi^2}{[L\sin(\pi\ell/L)]^{2}}+\frac{\pi^2}{18L^2}$,\citealp{calcont}
which has a larger exponent than the ones of $\alpha>$1. It is amazing
that a fit of those 8 points to Eq. (\ref{eq:fit}) give the following
values: $d_{1}=0.478,$ $f_{1}\times (L/\pi)^{2}=0.018$ and $p_{1}=2.05$.
These values are very close to the exact ones, i. e.: $d_{1}=0.4785$,
$f_{1}\times (L/\pi)^{2}=1/60=0.0166$ and $p_{1}=2$.\citealp{igloiising2008,calcont}

It is interesting to stress that the correction term of order $\ell^{-2}$
in the case of the IM does not come from the second term in (\ref{eq:diffentropyb})
with $\kappa_{Ising}=2\pi$, but from the third term with $\nu =2$.
This is due to the exact relations among the entropies of the IM and
the XX chain, and the fact that there is no oscillations in the last
model for $\alpha=1$, i. e. $g_{1}=0$.

\begin{table}
\begin{tabular}{cccc}
\multicolumn{1}{c}{} & $p_{2}$  & $p_{3}$  & $p_{4}$\tabularnewline
\hline
\hline 
 & 0.978  & 0.646  & 0.473\tabularnewline
 & (\emph{1.000})  & (\emph{0.6666})  & (\emph{0.5000})\tabularnewline
\hline
\hline 
 &  &  & \tabularnewline
\end{tabular}

\caption{\label{tab:Isingpalpha}The finite-size estimates of the exponents
$p_{\alpha}$ obtained from the fit of Eq. (\ref{eq:fit}) for the
Ising model with PBC and $L=96$. The results in parentheses are the
exact values.}

\end{table}

Let us now focus on the BCM. We show in the last column of Table
\ref{tab:BCMpalpha} the finite-size estimates of the central charge
$c$ for three values of $\gamma$ along the critical line. We got
these estimates by a simple fit of our data to Eq. (\ref{eq:entropyCFT})
for $\alpha=1$ . As we note, theses values of $\gamma$ satisfy $\gamma>\gamma_{\mbox{\scriptsize tr}}=0.41563$,
and the system belongs to the same universality class of critical behavior
as the Ising model with $c=1/2$. Following the same procedure done
in the IM, we fit our data to Eq. (\ref{eq:fit}) with the appropriated
values of $c$ {[}see Fig. \ref{fig1}(b){]}, in order to extract
the exponents $p_{\alpha}$. Some finite-size estimates of $p_{\alpha}$
are depicted in Table \ref{tab:BCMpalpha}. As observed in this table,
for $\gamma>\gamma_{\mbox{\scriptsize tr}}$, our results are consistent
with the exponent $p_{\alpha}=2/\alpha$, which is the same of the
IM. In the case $\alpha=1$, where the evaluations are more difficult
we got, in the BCM, around 10 points of $D_{\alpha}(L,\ell)$ satisfying
$|D_{1}(L,\ell)-D_{1}(L,\ell+1)|<10^{-5}$ for $\ell>5$, and the
fit to Eq. (\ref{eq:fit}) indicate that $p_{1}=2$. Since the energy
operator $X_{\epsilon}=1,$ this term could be produced either by
the first term in (\ref{eq:diffentropyb}) with $g_{1}\ne0$ but $\kappa=0$,
or by the second term in (\ref{eq:diffentropyb}) with $\nu =2$.
However, the exact result derived for the IM, that we expect to be
the same for any model on its universality class, indicate that $g_{1}=0$
and the dominant term in (\ref{eq:diffentropyb}) 
is the one with exponent $\nu =2$.
  Our estimates for this exponent $p_{1}=\nu $
are shown in Table \ref{tab:BCMpalpha}.

At the tricritical point $(\gamma_{tric},\delta_{tric})$, we were
not able to extract the exponents $p_{\alpha}$ due to the fact that
the $\alpha$-entropies are very sensitive to the coupling constants of the
model.
%, as illustrated in Fig. \ref{fig1}(c). In this figure, we show
%$S_{\alpha}(L,\ell)$ for $\gamma=1.1$ and two values of $\delta$:
%$\delta=\delta_{c}$ and $\delta=\delta_{c}+\epsilon$, where $\epsilon=0.5\times10^{-4}$.
We observed that a very small error $(\epsilon\sim 10^{-4})$ in
the critical coupling $\delta_{c}$ affects very little the value
of the entropy along the critical line. 
Typically, the entropy changes are
of the same magnitude of $\epsilon$. On the other hand, at the tricritical
point, a small change in the critical couplings
%, i. e. $(\gamma_{\mbox{\scriptsize tr}},i
%\delta_{\mbox{\scriptsize tr}})\rightarrow(\gamma_{\mbox{\scriptsize tr}}+\epsilon,\delta_{\mbox{\scriptsize tr}}+\epsilon)$,
produces a much larger effect. Due to this fact, since the precision
of the tricritical couplings are $\sim (10^{-5})$, which is of the same
order of magnitude as $D_{\alpha}$, we are not able to obtain a reasonable
estimate of  $p_{\alpha}$
at the tricritical point.

\begin{table}
\begin{tabular}{ccccccc}
\multicolumn{1}{c}{$\gamma$ } &  & $p_{2}$  & $p_{3}$  & $p_{4}$  & $\nu $  & $c$ \tabularnewline
\hline
\hline 
1.2 &  & 1.066  & 0.697  & 0.517  & 2.071 & 0.5000\tabularnewline
1.1 &  & 1.041  & 0.682  & 0.504  & 2.062  & 0.5001 \tabularnewline
0.7 & \hspace*{0.5cm} & 1.154  & 0.776  & 0.589  & 1.955 & 0.5015\tabularnewline
 &  & (1)  & (2/3)  & (1/2)  & (2)  & (1/2) \tabularnewline
\hline
\hline 
 & \multicolumn{1}{c}{} &  &  &  &  & \tabularnewline
\end{tabular}

\caption{\label{tab:BCMpalpha}The finite-size estimates of the exponents $p_{\alpha}$
acquired by the fit of Eq. (\ref{eq:fit}) for the BCM with PBC, $L=96$,
and some values of $\gamma$. In the last two columns we also present
the finite-size estimates of the conformal anomaly $c$, and the exponent
$p_{1}=\nu$ (see text).The values in parentheses are the
expected ones.}

\end{table}

\subsection{The three state Potts Model }

The quantum 3SPM is a quantum chain obtained by the $\tau$-continuum
limit of the two dimensional classical 3SPM.\citealp{revPotts} The
quantum Hamiltonian of the model can be written as 

\begin{equation}
H_{Potts}=-\sum_{i}(R_{i}R_{i+1}^{\dagger}+h.c.)+\lambda O_{i},\label{eq:potts}\end{equation}
where $R$ and $O$ are $3\times3$ matrices given by 

\[
R=\left(\begin{array}{ccc}
0 & 1 & 0\\
0 & 0 & 1\\
1 & 0 & 0\end{array}\right)\qquad O=\left(\begin{array}{ccc}
2 & 0 & 0\\
0 & -1 & 0\\
0 & 0 & -1\end{array}\right).\]

\begin{figure}[!t]
\psfrag{ell}{$\ell$}
\begin{centering}
%\psfrag{ell}{$\ell$}
\includegraphics[scale=0.24]{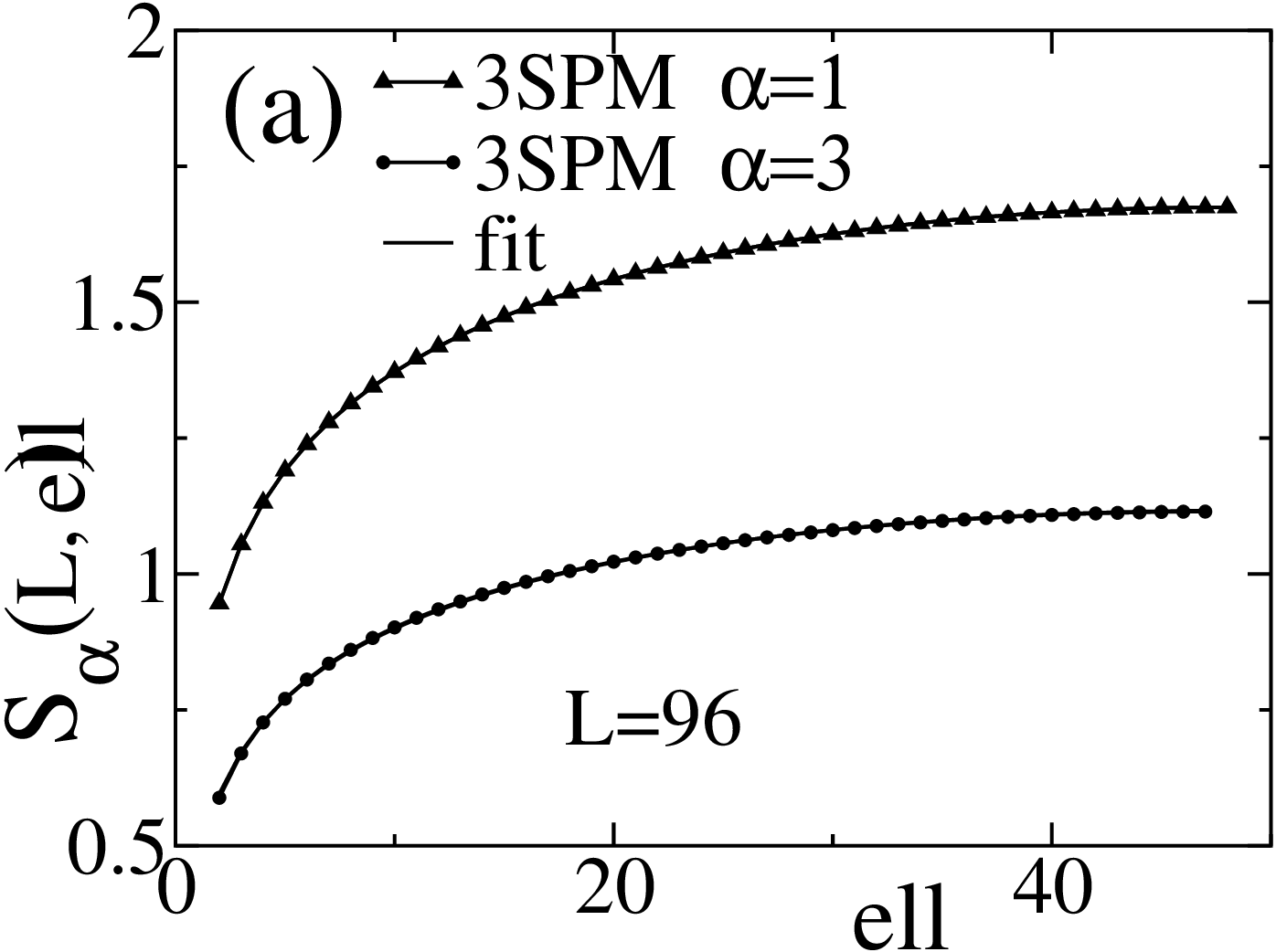}
\par\end{centering}

\begin{centering}
\includegraphics[scale=0.24]{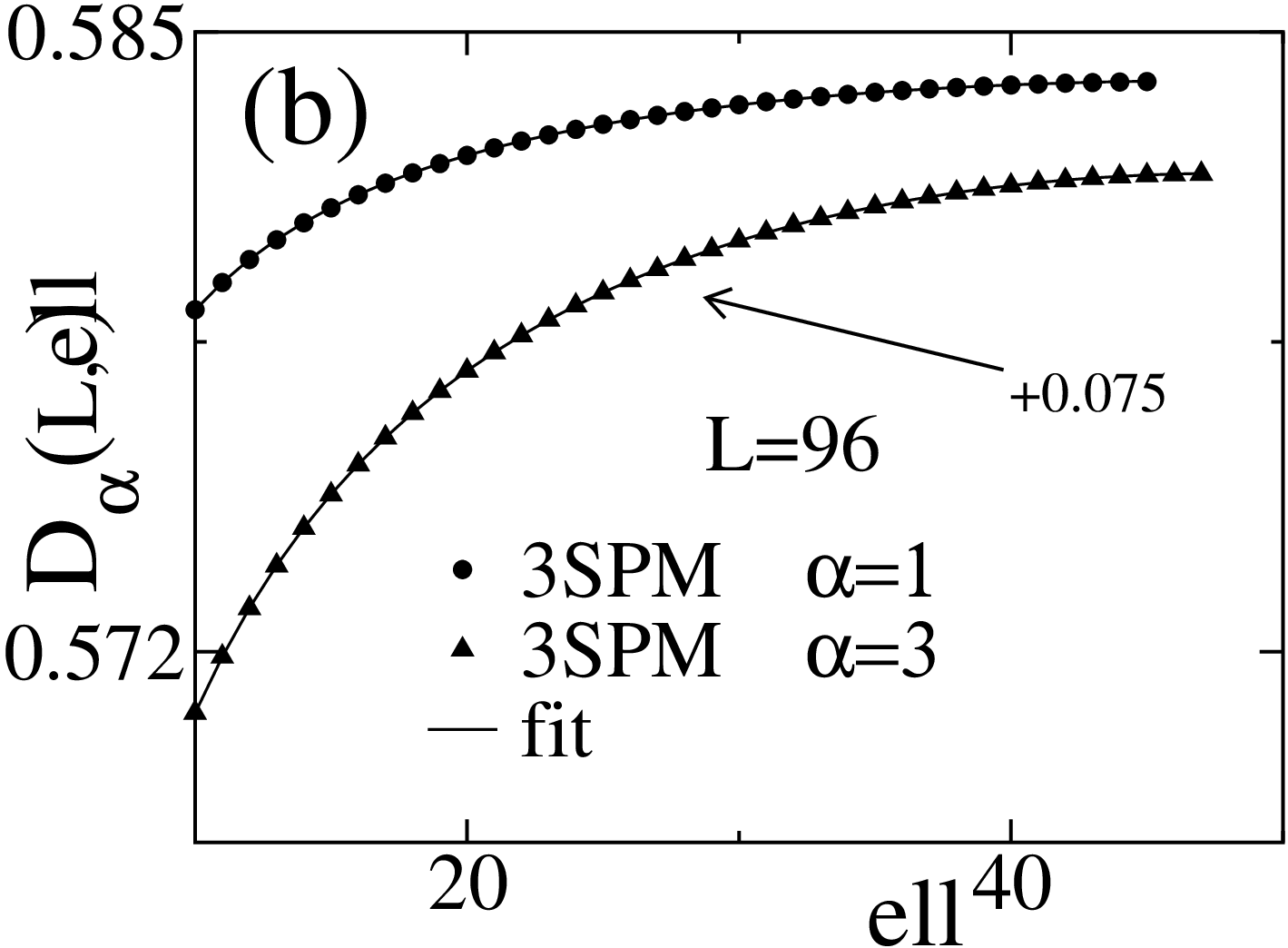}
\par\end{centering}

\caption{\label{fig2potts3} (Color online). Rényi entropy $S_{\alpha}(L,\ell)$
and the difference $D_{\alpha}(L,\ell)$ for the 3SPM with lattice
size $L=96$. The symbols are the numerical data and the solid lines
connect the fitted points (see text). (a) $S_{\alpha}(L,\ell)$ vs
$\ell$ for $\lambda=1$, and some values of $\alpha$ (see legend).
(b) $D_{\alpha}(L,\ell)$ vs $\ell$ for $\lambda=1$, $\alpha=2$
and $\alpha=3$.}

\end{figure}

As the IM, due to its self duality the 3SPM also has the critical
point at $\lambda_{c}=1$. However, the critical fluctuations of the
3SPM are described by a CFT with $c=4/5$. For PBC, as a consequence
of the modular invariance of the related two dimensional model defined
on a cylinder, the associated CFT is described in terms of 10 primary
operators.\citep{dimensions} Among these operators, we focus on the
most relevant operators, which may lead to the unusual corrections
to scaling in the entropy.

The first nonzero dimension ($X_{m}=2/15$) is associated to the order
parameter and the second one ($X_{\epsilon}=4/5$) is associated to
the energy operator. The lowest irrelevant operator responsible for
the finite-size corrections of the eigenenergies of the Hamiltonian
has dimension $X^{I}=14/5$.\citep{dimensions,chico1} 

Note that in the IM and the BCM, the leading corrections to scaling
in the $\alpha$-entropies with $\alpha>1$ are related to the dimension
of the energy operator $X_{\epsilon}$. For this reason, we expected
in the 3SPM that $p_{\alpha}=\frac{2X_{\epsilon}}{\alpha}=\frac{8}{5\alpha}$
for $\alpha>1$. Below we use the same procedure used in the IM and
the BCM to get the exponent $p_{\alpha}$.

In Fig. \ref{fig2potts3}(a), we present the Rényi entropy as a function
of $\ell$ for the 3SPM at the critical coupling $\lambda_{c}=1$
for a system with lattice size $L=96$. As observed in this figure,
we also do not see the parity oscillations, like the models considered
in the last section (which also have a discrete symmetry). We are
able, in this case, to obtain a nice fit of our data by considering
only the standard term of the CFT {[}Eq. (\ref{eq:entropyCFT}){]}.
The solid lines in these figures connect the fitted points. In this
case, we get $c=0.799$ (see Table \ref{tab:potts3palpha}) very close
to the expected value $c=4/5$.

Finally, we depicted in Table \ref{tab:potts3palpha} the exponents
$p_{\alpha}$ obtained through a fit of our data to Eq. (\ref{eq:fit}).
As we can see in this table our numerical results indicate that, for
$\alpha>1$, $p_{\alpha}=\frac{8}{5\alpha}$ confirming our prediction
that like in the IM and BCM the energy operator gives the most important
contribution in the conical singularities. In the case of
the von Neumann entropy, i. e. $\alpha=1$, our numerical results
indicate that the leading corrections are given by the 
 third term in (6) with the parameter $\nu =2$.
 As  happened in the Ising universality
class the  operator that gives the leading contribution for 
$\alpha >1$, that we believe to be the energy operator,
  does not contribute
for $\alpha=1$, i. e., $g_{1}=0$. Moreover, since the leading irrelevant
operator is $X^{I}=14/15$, we would expect a contribution of order
$\left[\sin(\pi\ell/L)\right]^{-8/15}$ in (\ref{eq:diffentropyb}),
which is also not present, i. e., $b_{1}=0$. This implies that both
the energy and the leading irrelevant operator ruling finite-size
corrections do not contribute, at least at leading order,
 to the usual von Neumann entropy. The
result $g_{1}=0$ found for all models discussed above indicate that,
probably, this is a general behavior for quantum critical chains. 

\begin{table}
\begin{tabular}{cccc}
$p_{2}$  & $p_{3}$  & $p_{4}$  & $c$\tabularnewline
\hline
\hline 
0.829  & 0.523  & 0.371  & 0.799\tabularnewline
(0.800)  & (0.5333)  & (0.400)  & (0.800)\tabularnewline
\hline
\hline 
 &  &  & \tabularnewline
\end{tabular}

\caption{\label{tab:potts3palpha}The finite-size estimates of the exponents
$p_{\alpha}$ acquired by fit of Eq. (\ref{eq:fit}) for the three
state Potts model for $L=96$. In the last column we also present
the finite-size estimates of the conformal anomaly $c$ (see text).
The results in parentheses are the expected values.}

\end{table}

\subsection{The spin-1 Fateev-Zamolodchikov quantum chain}

The spin-1 Fateev-Zamolodchikov quantum chain (FZQC) is an exact integrable
quantum chain whose Hamiltonian is given by \citep{spin1ZM}

\begin{align}
H_{FZ} & =\epsilon\sum_{j}\left\{ \mathbf{s}_{j}\cdot\mathbf{s}_{j+1}-\left(\mathbf{s}_{j}\cdot\mathbf{s}_{j+1}\right)^{2}\right.\nonumber \\
 & + 4\sin^{2}(\delta/2)\left(T_{j}^{\perp}T_{j}^{z}+h.c.\right)\nonumber \\
 & -\left.2\sin^{2}(\delta)\left[T_{j}^{z}-\left(T_{j}^{z}\right)^{2}+2\left(s_{j}^{z}\right)^{2}\right]\right\} ,\label{eq:ZM}\end{align}
where $\epsilon=\pm1$, $T_{j}^{z}=s_{j}^{z}s_{j+1}^{z}$, $T_{j}^{\perp}=s_{j}^{x}s_{j+1}^{x}+s_{j}^{y}s_{j+1}^{y}$,
and $s_{j}^{x}$, $s_{j}^{y}$, $s_{j}^{z}$ are the spin-1 $SU(2)$ operators. 

\begin{figure}[!t]
\psfrag{ell}{$\ell$}
\begin{centering}
%\psfrag{ell}{$\ell$}
\includegraphics[scale=0.24]{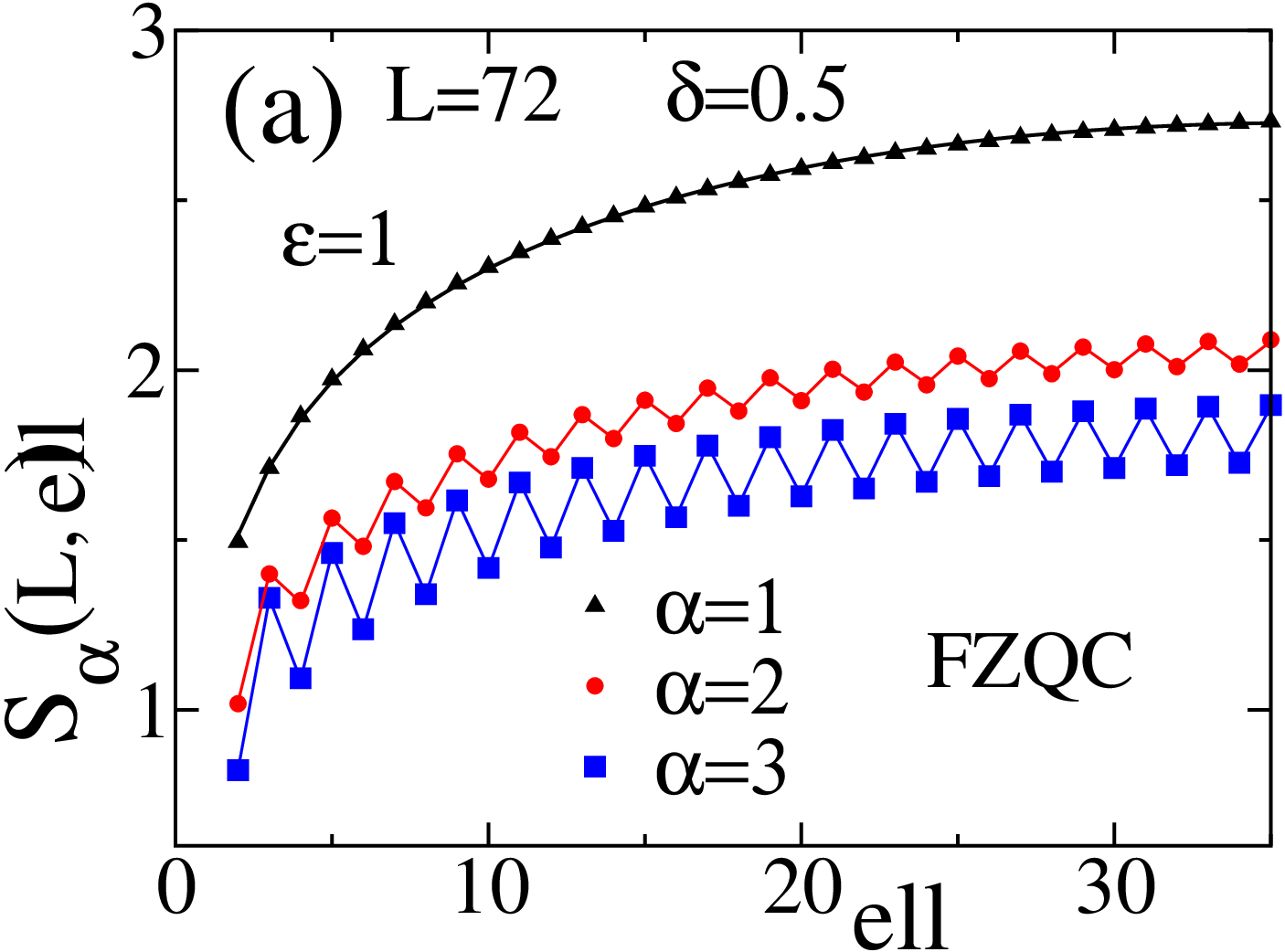}
\par\end{centering}

\begin{centering}
\includegraphics[scale=0.24]{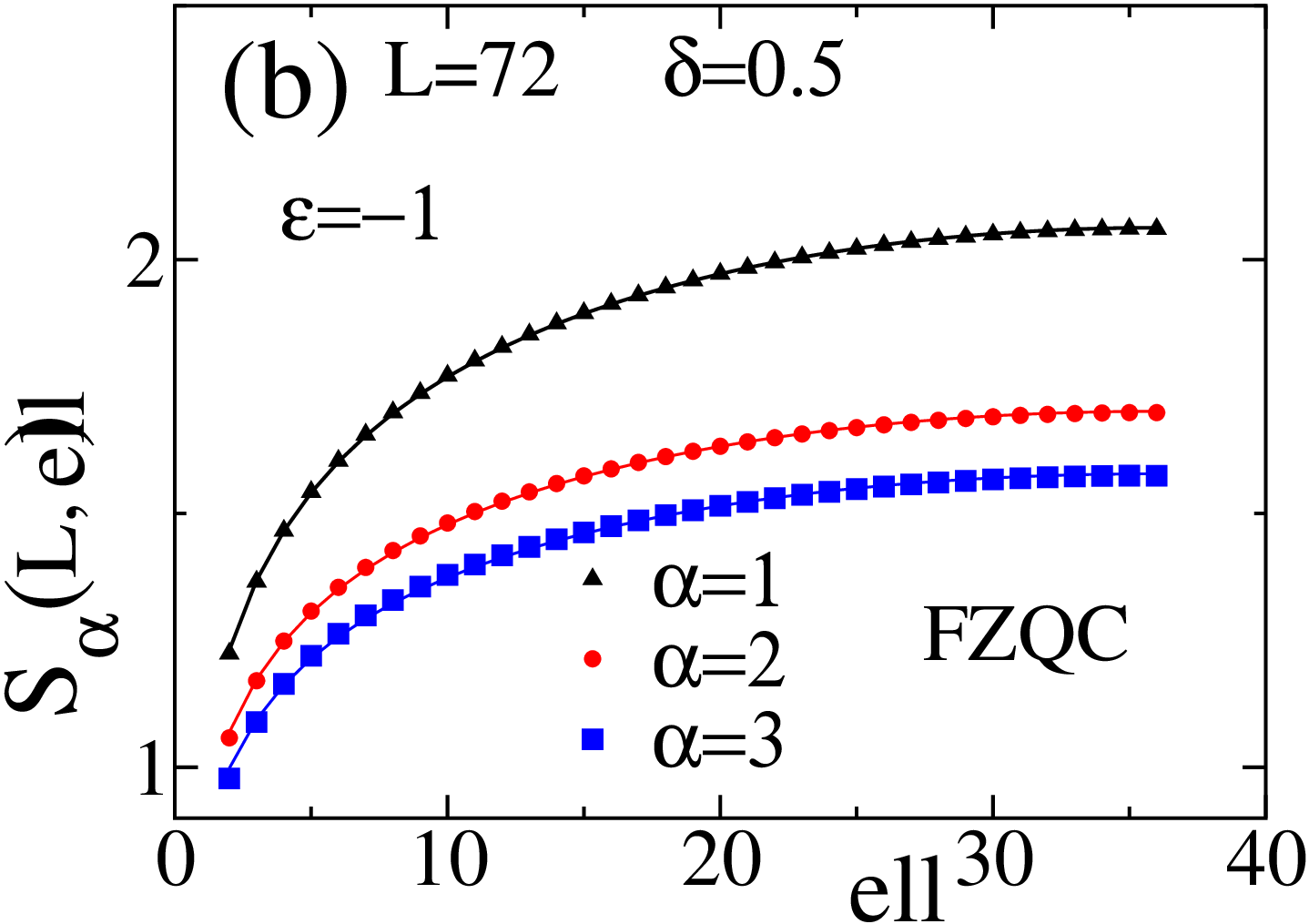}
\par\end{centering}

\begin{centering}
\includegraphics[scale=0.24]{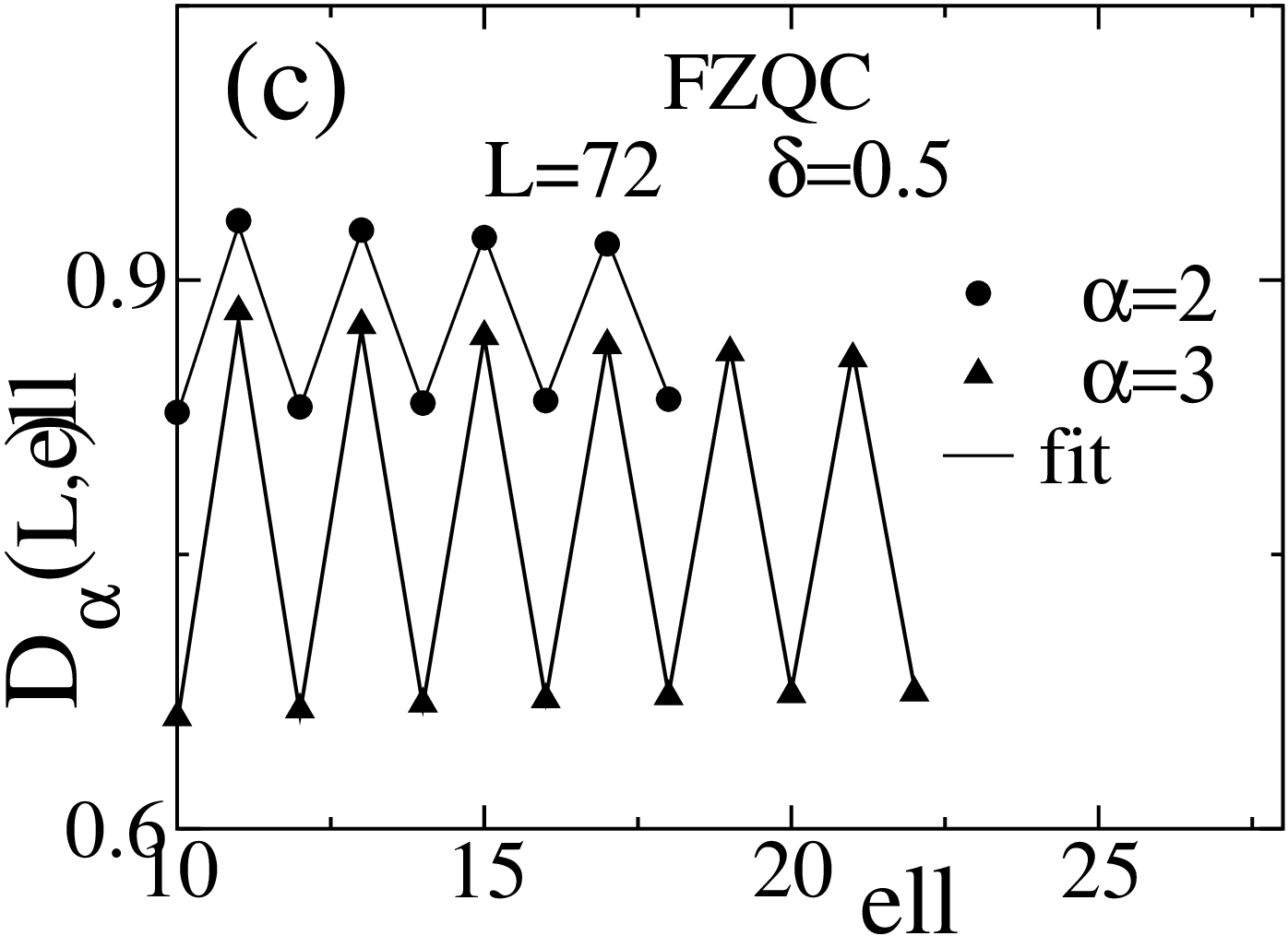}
\par\end{centering}

\caption{\label{fig3ZF} (Color online). Results of the Rényi entropy $S_{\alpha}(L,\ell)$
and the difference $D_{\alpha}(L,\ell)$ for the spin-1 Fateev-Zamolodchikov
quantum chain with PBC and size $L=72$. The symbols are the numerical
data and the solid lines connect the fitted points (see text). (a)
$S_{\alpha}(L,\ell)$ vs $\ell$ for $\epsilon=1$, $\delta=0.5$,
and some values of $\alpha$ (see legend). (b) Same as (a) but for
$\epsilon=-1$. (c) $D_{\alpha}(L,\ell)$ vs $\ell$ for $\epsilon=1$,
$\alpha=2$ and $\alpha=3$ at $\delta=0.5$. }

\end{figure}

This Hamiltonian has a $U(1)$ symmetry, having the $z$-magnetization
as a good quantum number. The antiferromagnetic ($\epsilon=1$) and
ferromagnetic ($\epsilon=-1$) models show a critical phase with continuously
varying exponents for $0\leq\delta\leq\pi/2$. The antiferromagnetic
and ferromagnetic models are governed by a CFT with central charge
$c=3/2$ and $c=1$, respectively.\citep{PhysRevLett.61.1529,gasSu2,spin1Zmchico,spin1ZMVlad,PhysRevLett.63.708}
The CFT in the $c=3/2$ case is described in terms of composite fields
formed by Gaussian and Ising operators.\citep{PhysRevLett.61.1529,gasSu2,spin1Zmchico,spin1ZMVlad}
The energy operator, which we believe is responsible for the leading
contributions due to the conical singularities, has dimension $X_{\epsilon}=\frac{\pi}{4(\pi-2\delta)}+\frac{1}{8}$
and the leading irrelevant operator responsible for the finite-size
corrections of the eigenenergies has dimension $X^{I}=\frac{\pi}{\pi-2\delta}+1$.\citep{PhysRevLett.61.1529,spin1Zmchico} 

In the case $c=1$, we have a Gaussian CFT or a standard Luttinger
liquid, whose Luttinger parameter changes continuously along the critical
line. In this case, the dimension of the energy operator is $X_{\epsilon}=\frac{\pi}{2\delta}$
and the leading irrelevant operator responsible for the corrections
to scaling has dimension $X^{I}=\frac{2\pi}{\delta}$.\citep{PhysRevLett.63.708}

In Figs. \ref{fig3ZF}(a)-(b), we present the Rényi entropy $S_{\alpha}(L,\ell)$
for spin-1 FZQC with, PBC, $\delta=0.5$, and size $L=72$. Notice
that for $\epsilon=1$, differently from the Ising and the Blume-Capel
models, the $\alpha-$Rényi entropies oscillate for $\alpha>1$, which
is a signature of the 
oscillating spin-spin correlations of the quantum chain.
The absence of oscillations for $\alpha=1$ also suggests that like
the other models we have studied in previous subsections $g_{1}=0$
in (\ref{eq:diffentropyb}). Let us focus, first, on the case $\alpha=1$.
In this case, we get very nice fit of our numerical data (black triangles)
to the standard conformal field term $S_{\alpha}^{CFT}$ {[}Eq. (\ref{eq:entropyCFT}){]}.
The fit gives us $c=1.491$ ($c=1.02)$ for $\epsilon=1$ ($\epsilon=-1)$,
which is very close to the exact value $c_{exact}=3/2$ ($c_{exact}=1$).
Similar estimates of $c$ were also obtained by considering $\alpha=2$
or $\alpha=3$. Note that the central charge $c$ of the FZQC is larger
than the ones of Ising and Blume-Capel models. This means that the
subsystem $\mathcal{A}$ is more entangled with the subsystem $\mathcal{B}$
in the FZQC, as compared with the critical Ising or Blume-Capel models.
Due to this fact, we have to keep a much larger number of states $m$
(typically around $m\sim2000-3000$) for the FZQC, in order to get
results with some similar accuracy.

\begin{table}
\begin{tabular}{cc|ccc}
\multicolumn{1}{c}{} & \multicolumn{1}{c|}{$\delta$} & $p_{2}$  & $p_{3}$  & $p_{4}$\tabularnewline
\hline
\hline 
 & 0.25  & 0.413  & 0.286  & 0.215\tabularnewline
 &  & (\emph{0.4223})  & (\emph{0.2815})  & (\emph{0.2112})\tabularnewline
 &  &  &  & \tabularnewline
 & 0.5  & 0.432  & 0.304  & 0.231\tabularnewline
 &  & (\emph{0.4917})  & (\emph{0.3278})  & (\emph{0.2459})\tabularnewline
 &  &  &  & \tabularnewline
 & 1.0  & 0.778  & 0.541  & 0.402\tabularnewline
 &  & (\emph{0.8129})  & (\emph{0.5419})  & (\emph{0.4064})\tabularnewline
\hline
\hline 
 & \multicolumn{1}{c|}{} &  &  & \tabularnewline
\end{tabular}\caption{\label{tab:palphaZM}The finite-size estimates of the exponents $p_{\alpha}$
acquired by fit of Eq. (\ref{eq:fit}) for the spin-1 Fateev-Zamolodchikov
quantum chain with PBC, $L=72$ and some values of $\delta$. The
values in parentheses are the expected ones (see text).}

\end{table}

Now let us consider the antiferromagnetic model $\epsilon=1$ and
$\alpha>1$. In this case, it is not possible to fit the numerical
data if we consider only the standard CFT term $S_{\alpha}^{CFT}$
{[}Eq. (\ref{eq:entropyCFT}){]}. Relative nice fits are obtained
only if we consider besides the CFT term, the oscillatory correction
term {[}Eq. (\ref{eq:entropyUnusual}){]} with $\kappa=\pi/2$. The
solid lines in Fig. \ref{fig3ZF}(a) are the fits to Eq. (\ref{eq:entropyb}).
For instance, for $\alpha=2$ ($\alpha=3$) we get $c=1.48$ ($c=1.45)$
and $p_{2}=0.54$ ($p_{3}=0.36)$. 

Since we are confident that $c=3/2$ for the spin-1 FZQC, we set $c=3/2$
in Eq. (\ref{eq:entropyCFT}), as well $\kappa=\pi/2$ in the function $D_{\alpha}(L,\ell)$
{[}Eq. (\ref{eq:diffentropyb}){]}, in order to evaluate the exponent
$p_{\alpha}$ in the same way we did in the Ising, Blume-Capel and
three state Potts models. In Fig. \ref{fig3ZF}(b), we show $D_{\alpha}(L,\ell)$,
as function of $\ell$, for the spin-1 FZQC with PBC, $\epsilon=1$,
$\delta=0.5$, $L=72$ and three values of $\alpha$. Even keeping
$m=3000$ states in the DMRG, it is very hard to get high precision in
the values of the entropy for the spin-1 FZQC, as we already mention
before. Although the truncation errors are around $\sim10^{-8}$, comparison
of the entropy results with different values of $m$, indicate that the
errors in the entropy are around $\sim10^{-3}$. For this reason, we
discarded the points that $|D_{\alpha}(L,\ell)-D_{\alpha}(L,\ell+2)|<10^{-3}$,
as well as the first 10 sites. The solid lines in this figure connect
the fitted points. The finite-size estimates of $p_{\alpha}$ in (\ref{eq:fit}),
obtained by this fitting procedure, are depicted in Table \ref{tab:palphaZM}.
The results presented in this table indicate that $p_{\alpha}=2X_{\epsilon}/{\alpha}$,
where $X_{\epsilon}=\frac{\pi}{4(\pi-2\delta)}+\frac{1}{8}$ is the
dimension of the \textit{energy operator} of the model. This means
that, like the other quantum chains, the energy operator gives the
most important contribution in the conical singularities for $\alpha>1$.
In the case where $\alpha=1$, like the other models, this contribution
seems to be absent, i. e., $g_{1}=0$ {[}see Fig.~\ref{fig3ZF}(a){]}.
In this case, we were not able to estimate numerically the power of
the leading correction, since we only have three points satisfying
$|D_{1}(L,\ell)-D_{1}(L,\ell+2)|<10^{-3}$.

In the ferromagnetic case ($\epsilon=-1$) although we expect oscillations
to be present in the $\alpha$-entropies with $\alpha>1$, we could
not see them numerically {[}see Fig.~\ref{fig3ZF}(b){]}. This is
consistent with the conjecture that the oscillations are ruled by
the energy operator. In this case the amplitude of the oscillations
will decay like $\ell^{-p_{\alpha}}$, $p_{\alpha}=\frac{2X{_{\epsilon}}}{\alpha}=\frac{\pi}{\delta\alpha}$.
The region where we would better see the oscillations would be for
$\delta\approx\frac{\pi}{2}$, where $p_{\alpha}$ has the smaller
values. However the sound velocity, which is given by \citealp{PhysRevLett.63.708}
$v_{s}=\pi\sin(2\delta)/(2\pi-2\delta)$ is close to zero in this
region. This makes the convergence quite slow in the DMRG, due to
the small size of the energy mass gaps. To avoid this problem and
get enough precision, we considered values of $\delta$ in other regions,
like $\delta=1/2$ for example {[}see Fig.~\ref{fig3ZF}(b){]}. However,
in this case, the expected value of $p_{\alpha}=2\pi/\alpha$ is for
$\alpha=3$, $p_{\alpha}\approx2.1$. This gives a strong decay, large enough 
to forbid, within the numerical precision we have, the observation
of the oscillations for $\ell>10$.

\section{Conclusions}

Most of the critical quantum chains are conformal invariant. This
symmetry implies that the mass gap amplitudes of these critical chains,
in a finite lattice, are related to the conformal anomaly and critical
exponents that label the particular universality class of critical
behavior.\citep{cardydomb} Due to these relations, the most frequently used
method to extract the conformal anomaly and critical exponents comes
from the finite-size scaling of the \textit{eigenenergies} of the
critical quantum chains.

On the other hand, the conformal anomaly can also be calculated from
the Rényi entanglement entropies of the ground state of finite quantum
chains,\citep{cold,prlkorepin,cardyentan,cvidal} a property related
to the \textit{eigenfunctions} instead of the \textit{eigenenergies}.
These results raised the question about the possibility to extract
all the critical exponents from a complete finite-size study of the
Rényi entropies of the critical quantum chains. This question has
also practical implications, since in DMRG calculations the entanglement
entropies are much simpler to calculate, as compared with the mass
gap energies. Unfortunately the leading terms of the entanglement
Rényi entropies of the low-lying excited states are the same as that
of the ground state. \citealp{chicoentropy,chicoentropyext2} The
critical exponents only appear in the finite-size corrections of these
entropies.\citealp{entropyosc,cardyosc} In the finite-size study
of the energy gaps the exponents are given by the leading terms and
we know precisely the correspondence between the mass gap amplitudes
and the dimensions of the underlying CFT. In the finite-size study
of the entanglement entropy, since the exponents are given not by
the leading term, but by the finite-size corrections, we do not have
such exact correspondence. It is known from conformal invariance\citealp{cardyosc}
that these corrections are ruled either by relevant operators, due
to the conical singularities in the conformal mapping, or by irrelevant
operators. However, it is not known what are those operators, or equivalently
what should be the critical exponent that appears in the leading finite-size
corrections of the entanglement entropies in general models. The results
presented in this paper, together with known results induce us to
announce some conjectures about the operators governing the finite-size
corrections of the Rényi entropies of critical quantum chains with
PBC. We are going to state them separately

\textit{(a)} Entropy oscillations in the $\alpha$-Rényi entropies
of critical quantum chains only appear for $\alpha>1$ and for models
having at least one $U(1)$ symmetry, like the spin-$s$ Heisenberg,
$t-J$ and Hubbard models. Models possessing only discrete symmetries
like the Ising, Blume-Capel and Potts models show no oscillations
for any value of $\alpha$.

\textit{(b)} For any critical quantum chain the amplitude of the leading
correction of the $\alpha$-Rényi entropy with $\alpha>1$ has a universal
power decay $p_{\alpha}=2X_{\epsilon}/{\alpha}$, where $X_{\epsilon}$
is the dimension of the \textit{energy operator} of the model.
We  stress that these results are expected  only for the entanglement 
entropy of a single interval. 
%in the cases w here we split
% the system in two continuous subsystems. 
In the case of two disjoint intervals, the results of Ref. \onlinecite{albatwo} 
%we split the system in 4 subsyst
%(two separated subsystems) the results of Ref. \onlinecite{albatwo} 
indicate that instead of the energy operator the leading contributions 
comes from nonlocal spinor operators.

\begin{table}
\begin{tabular}{cccccc}
$\Delta$  &  & $M=0$  & $M=1/6$  & $M=1/4$  & $M=3/10$\tabularnewline
\hline
$0.5$  &  & 1.9  & 2.0  & 2.0  & 2.1\tabularnewline
$\sqrt{2}/2$  & \hspace*{.5cm} & --- & 2.0  & 2.0  & 2.1\tabularnewline
$0.9980$  &  & ---  & 1.98  & 2.0  & 2.1\tabularnewline
$2.0$  &  & ----  & 1.97  & 2.0 & ----\tabularnewline
\hline
\hline 
 &  &  &  &  & \tabularnewline
\end{tabular}

\caption{\label{tab:XXZ}The exponent $\nu$ of the finite-size corrections of the
von Neumann entropy for the spin-1/2 XXZ chain with PBC, and some values of $\Delta$
and magnetization per site $M$. We extracted the exponent, as we
did for the other models, considering systems with size $L=96$. For
value of anisotropy close to the isotropic antiferromagnetic point
($\Delta=1)$ and $M=0$, we were not able to obtain the exponent,
mainly due to the large finite-size corrections.}

\end{table}

The amplitudes of the leading corrections for $\alpha=1$, have a
quite distinct behavior from the $\alpha>1$ cases. For all models
we considered these amplitudes have a power law decay $\ell^{-\nu}$,
where $\nu =p_1=2$.
Since these corrections for the $\alpha=1$ Rényi entropy (von Neumann
entropy) are not known in the case of the spin-1/2 XXZ chain {[}anisotropy
$\Delta${]}, we present in Table \ref{tab:XXZ} those corrections
for some values of anisotropy and magnetization per site $M$. We
clearly see, that also for the XXZ quantum chain the leading finite-size
correction always decay as $\ell^{-2}$. In the case of the XX quantum
chain, where $\nu =2$ is an exact result,\citealp{xxPBCh}
or in the XXZ chain we cannot identify the operator responsible for
such finite-size correction. 
%In these cases, there exist two operators
%with dimension $\bar{X}_{R}=2$: the marginal operator that governs
%the continuous change of the exponents as we change the $z$-anisotropy,
%and the \textit{energy-momentum stress tensor} (dimension $\bar{X}_{R}=0+2=2$).
%However for the Ising, Blume-Capel and three-state Potts models there
%exists no additional operator with dimension $\bar{X}_{R}=2$, beyond
%the \textit{energy-momentum stress tensor}, that is always present
%in a conformal invariant quantum chain. 
All these results indicate
the following conjecture.

\textit{(c)} The leading finite-size corrections of the $\alpha=1$
Rényi entropy, or the von Neumann entropy, of any quantum chain decays
as $\ell^{-2}$. 
This is the main contribution coming from the conical singularities 
in the conformal mapping, but we do not know what operators produce such 
universal correction.

%This correction, for general models, are produced
%by the \textit{energy-momentum stress tensor operator} at the conical
%singularities, that happens in the conformal mapping that gives the
%reduced transfer matrices of the quantum chains.

We expect that the above conjectures can be confirmed for other models
and we hope they can be understood by using the general properties
of the underlying CFT ruling the long-distance physics of quantum
chains.
\begin{acknowledgments}
%We thank F. Essler, R. G. Pereira, V. Rittenberg and M. J. Martins for 
%useful discussions. 
We are in debt to P. Calabrese, F. Essler and R. G. Pereira for  
discussions and a careful  reading of the manuscript. We also thank M. J. Martins and
V. Rittenberg for useful discussions.
This research was supported by the Brazilian agencies FAPEMIG, FAPESP,
and CNPq. 
\end{acknowledgments}
\bibliographystyle{apsrev4-1}
%\addcontentsline{toc}{section}{\refname}\bibliography{/home/jcxavier/FILES/textosJ2011/refs_rev4}

\begin{thebibliography}{10}
\bibitem{commCFT} There exists in the underlying CFT describing 
the XXZ quantum chain two primary operators with the same dimension $X^{con}=K$.
 The energy operator actually is a combination of these two operators.
\end{thebibliography}
%Merlin.mbs v4.21 2009-07-09.
%

\end{document}